\documentclass[letter, traditabstract]{aa}
\usepackage{graphicx}
\usepackage{epsfig}
\usepackage{color} 
\usepackage{txfonts}
\usepackage{natbib}
\bibpunct{(}{)}{;}{a}{}{,} 

\begin{document}

\title{Clumpy streams in a smooth dark halo: the case of Palomar 5}
\titlerunning{Substructures in Pal 5's tidal tails}

\author{Alessandra Mastrobuono-Battisti\inst{1}, Paola Di Matteo\inst{2}, Marco Montuori\inst{3}, Misha Haywood\inst{2}}

\authorrunning{Mastrobuono-Battisti et al.}

\institute{Dipartimento di Fisica, Universit$\mathrm{\grave{a}}$ di Roma La Sapienza, Piazzale Aldo Moro 2, 00185 Rome, Italy \\
\email{alessandra.mastrobuonobattisti@uniroma1.it}
\and
GEPI, Observatoire de Paris, CNRS, Universit\'e
  Paris Diderot, 5 place Jules Janssen, 92190 Meudon, France
  \and
  ISC-CNR \& Dipartimento di Fisica, Universit$\mathrm{\grave{a}}$ di Roma La Sapienza, Piazzale Aldo Moro 2, 00185 Rome, Italy 
}

\date{Accepted, Received}

\abstract{
By means of   direct N-body simulations and simplified numerical models, we study  the formation and characteristics of the tidal tails around Palomar 5, along its orbit in the Milky Way potential. 
Unlike previous findings, we are able to reproduce the  substructures observed in the stellar streams of this cluster, without including any lumpiness in the dark matter halo.
We show that  overdensities similar to those observed in Palomar 5 can be reproduced by  the epicyclic motion of stars along its tails,  i.e. a simple local accumulation of orbits of stars 
that escaped from the cluster with very similar positions and velocities. This process is able to form stellar clumps at distances of several kiloparsecs from the cluster, so it is not a phenomenon confined to the inner part of Palomar 5's tails, as previously suggested. 
Our models  can reproduce the density contrast between the clumps and the surrounding tails found in the observed streams,  without including any
lumpiness in the dark halo, suggesting new upper limits on its granularity.
}
 

\maketitle

\section{Introduction}

At a distance of $23.2$~kpc from the Sun and of $16.6$~kpc above the Galactic plane, Palomar 5 
{ (Pal5)}
 is a halo globular cluster (GC) characterized by very low total mass, $M_{tot}$ = 5 $\times 10^3 {\rm M}_\odot$, a large core radius, $r_c=24.3$~pc, and low central concentration, $c$=0.66 \citep[see][]{oden02}. 
In the 
{ past}
 decade, a number of observational studies  \citep{oden01, oden03, grill06, jordi10} have shown 
{ there are}
 two massive  tidal tails emanating from opposite side of the cluster, made of stars escaped from the system 
 { owing}
  to the action of the Galactic tidal field. With an overall detected extension of $22^{\circ}$  on the sky, corresponding to a projected spatial length of  more than $10$~kpc, Pal 5's tails are so elongated and massive 
  { that they} 
  contain more stellar mass than the one currently estimated to be in the system itself \citep{oden03}. 
The resulting two tails are very narrow, with a projected, almost constant width of $120$~pc \citep{oden03}, comparable { to} the tidal radius of the cluster and kinematically cold \citep{oden09}, in agreement with what can be expected from the disruption of a GC \citep{capuzzo05}. Together with the tidal   associated to NGC 5466 \citep{grilljohn06}, Pal 5's tidal tails represent, so far, the most impressive example of the action of the Galactic tidal field on a GC.
The large spatial extension of Pal 5's tails naturally permitted study { of} their { fine structure}.  \\
\citet{oden03} were the first to show evidence of inhomogeneities in Pal 5's streams, characterized by stellar density gaps (underdense regions) and clumps (overdense regions), particularly visible in the trailing stream, with the most evident and massive overdensities found between $100$ and $120$~arcmin from the cluster center. The presence of density fluctuations along the stream { has been} successively confirmed by  \citet{grill06} and \citet{jordi10}{, and the }
nature of these clumps has been long debated. Already before the observations of Pal 5's tidal stream,  \citet{combes99} showed the presence of small clumps  in the tidal tails of simulated GCs, and attributed their formation to strong gravitational shocks suffered by the system at each disk crossing.  When analyzing a set of simulations of GCs orbiting in a triaxial potential, 
\citet{capuzzo05} and  \citet{dimatteo05}, proposed an alternative explanation,  suggesting for the first time that the clumps observed in their simulations 
were { instead} kinematic effects, related to a local deceleration in the motion of stars along the tails. \citet{kupper08},  \citet{just09}, and more recently  \citet{kupper10, kupper12} and  \citet{lane12} { show} that these kinematic effects and local decelerations correspond to ``epicyclic cusps'',  where stars escaping from the cluster slow down in their epicyclic motion, as seen in the cluster reference frame. Because stars cross the tidal boundary with similar position and velocities, the location of such epicyclic cusps is similar for most escapers, generating visible overdensities. \\ While all these works have successfully reproduced the presence of stellar substructures in tidal streams of GCs orbiting in an external potential, until now no simulation or model has been able to reproduce the only stellar clumps robustly observed in the tails of a galactic GC: those of Pal 5\footnote{Clumps have been observed also along the GD-1 stellar stream, whose progenitor was  presumably a GC, see  \citet{kopo10}.}.
In the most complete numerical study of Pal 5's stellar streams realized so far, \citet{dehnen04}
have not succeeded in reproducing the clumpy nature of the stream, even after adopting a realistic Galactic model, and they propose
 that the observed clumps may be 
 the effect of Galactic substructures not accounted { to} in their simulations, { such} as,  dark matter subhalos.
 \citet{yoon11} have recently discussed,  by means of simplified numerical models, the absence of clumpy substructures in Pal 5's tails when the cluster is evolved in a smooth dark matter halo, { instead} showing that stellar  overdensities and  gaps between them can be generated if the gravitational effect of dark subhalos on the streams is taken into account. \citet{carlberg12} { accounts} for the possibility that epicyclic motions could explain the density variations observed in the inner part of the tails, but excluded {the possibility} they could reproduce those  beyond about 1 kpc (2.5$^\circ$) from the cluster center. From the density of underdense regions in observed narrow streams (Pal 5, among others), { he estimates} a total halo population of about $10^5$ dark matter subhalos with masses greater than $10^5 {\rm M}_\odot$.   \\
The results we present in this Letter  reproduce, for the first time, the inhomogeneous properties of Pal 5's streams without invoking any lumpy dark halo. As { described} in the following sections,  epicyclic pileup of stars can be  responsible { for} generating these substructures, even at significant distances from the cluster. We also exclude either gravitational instabilities or episodic mass loss from the cluster as processes that may explain their formation. \\

\section{Models and initial conditions}\label{method}

\subsection{N-body simulations and restricted three-body methods} 
{ We have studied} the formation and characteristics of Pal 5's tidal tails by means of N-body, direct summation simulations, and simplified numerical models.
To run the simulations we used a modified version of NBSymple \citep{cmm11}, a high performance  direct $N$-body code implemented on a hybrid CPU+GPU platform by means of a double-parallelization on CPUs and on the hosted graphic 
processing units (GPUs). 
The effect of the external galactic field is taken into account { by} using an analytical representation of its gravitational potential (see \S \ref{gal_mod_tails}). 
We { ran several} simulations, varying, among other parameters, the value of the softening length adopted to smooth the mutual interactions. In this Letter we present the results of a simulation with a softening length $\varepsilon=0.03$~pc, but we want to emphasize that in all these simulations, the presence of substructures was detected.
The time integration of the particles trajectories { was} done using the second-order leapfrog method, with a time step $\Delta t=\sqrt{\varepsilon^3/(GM)}$, where $M$ is the total mass of the cluster. To fully exploit the GPUs power the force { was computed} in emulated double precisionWith these choices, the average relative error per time step { was} less than $10^{-13}$ for the cluster both isolated or in the Galactic potential.

The result of N-body simulations { were} compared to simplified, restricted three-body models, where  we { used} the same initial conditions adopted to run the simulation. However, at each time step, instead of evaluating the mutual interactions among all the $N$ particles in the system, we evaluated the interaction with the global gravitational potential of the cluster and with the Galactic potential
{ for each particle}. This approach allows the main characteristics of stellar streams to { be reproduced} (orientation, spatial distribution and extension, presence of overdense and underdense regions) without { having to make} CPU-time consuming N-body simulations. It represents, of course, a simplified model that does not take the evolution of the internal parameters of the system (its mass and concentration) into account and thus must be taken as a first-order approach in the study of tidal streams.

Finally, { similar} to \citet{lane12}, we { produced} streaklines that allow for a simple visualization of the spatial distribution, at the present time, of stars that have escaped from Pal 5 in the last Gyrs.
To do so, we { integrated} the orbit of a point mass representing Pal 5 barycenter, and, at each time step, { released} two particles at distance $r_\mathrm{max}=120$ pc from it, along the instantaneous Galactic center and anticenter directions, and with a velocity equal to that of the cluster. The motion of these escapers is  integrated in time, taking both the effects of the GC and Galactic potential into account. 

\begin{figure}
\centering
\includegraphics[trim=3.5cm 2.5cm 2.5cm 2.5cm, clip=true, width=0.35\textwidth, angle=270]{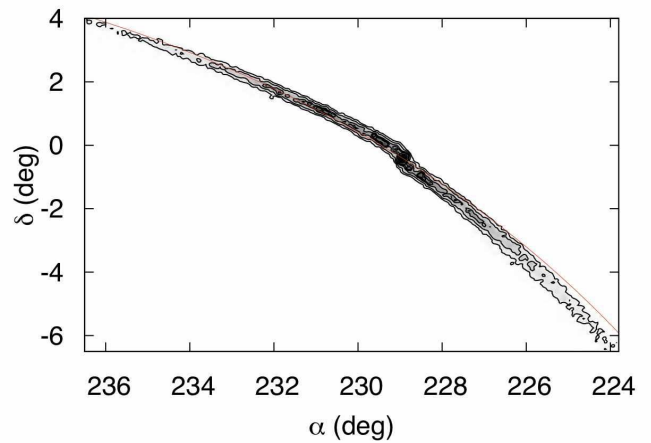}\\
\caption{Map of the surface density of stars of the simulated cluster in the plane of the sky. The solid red
line represents the local orbit.}
\label{isoall}
\end{figure}

\subsection{Globular cluster and Galaxy models}\label{gal_mod_tails} 
The Galactic potential is from  \citet{AS91}, whose model consists of a three-component system: a spherical central bulge
and a flattened disk, both { in the} \citet{MN75} form, plus a massive \emph{smooth} spherical halo.  We { did not include} a barred potential, since its effect on the orbit of Pal 5 has been shown to be negligible  \citep{AMP06}.

For the GC we adopted the parameters of the ``model A'' described by \citet{dehnen04}, because, according to the authors,  this model is the one that best fits the stellar number density of Pal 5 at the end of their simulations \citep[see Fig.~12,][]{dehnen04}. It consists of a { single-mass} King model \citep{Ki66} with tidal radius $r_t=56~$pc, $W_0=2.75$, and $M_0=2\times 10^4~{\rm M}_\odot$. To generate  positions and velocities of the $N=15360$ particles that represent the cluster we used NEMO \citep{Te95}.

\begin{figure}
\centering
\includegraphics[trim=4.6cm 1cm 1cm 0cm, clip=true, width=0.32\textwidth, angle=270]{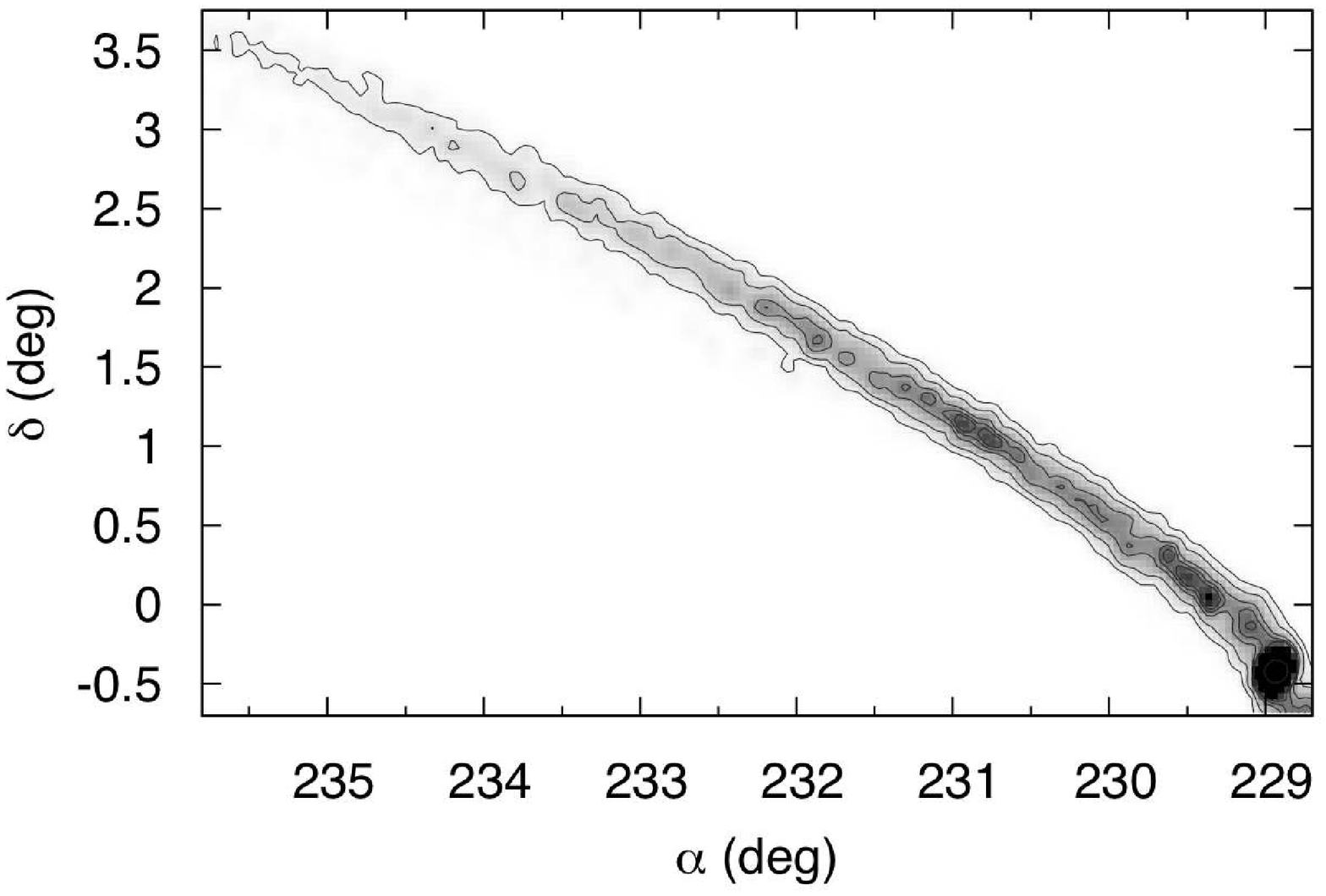}\\
\vspace{-2.cm}\hspace{-0.26cm}
\includegraphics[width=0.311\textwidth, angle=270]{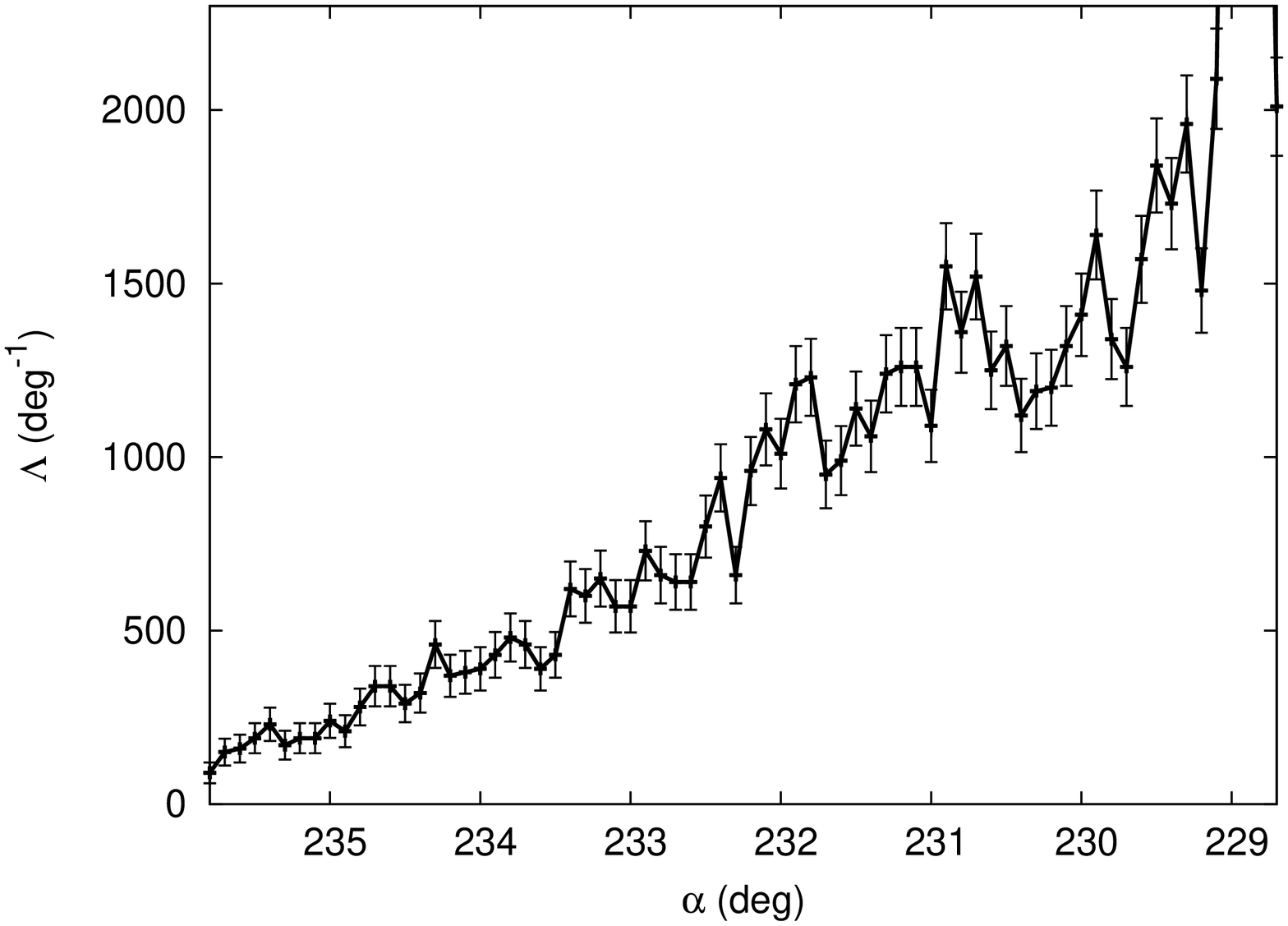}
\vspace{0.2cm}
\caption{Upper panel: The trailing  tail at the end of the N-body simulation. The contour lines in the tails refer to $740$, 
$2200$, $4800$, $6300$ and $7400$~M$_\odot/\rm{deg}^2$.  Bottom panel: Linear density of the trailing tail shown in the upper panel, as a function of the right ascension $\alpha$.  The cluster is on the right of the panel. Error bars are Poisson errors.
}  
\label{all}
\end{figure}
For the orbital initial conditions, we { first} integrated a test particle with Pal 5's present position and velocity backward in time for the same time interval as chosen by \citet{dehnen04}, i.e. $2.95$~Gyr, in the Galactic potential,  and obtained the initial conditions of the cluster barycenter. Then we translated the positions and velocities of the { single-mass} King model  previously described to coincide with those of the cluster barycenter $2.95$~Gyr ago, and  we integrated the whole cluster $2.95$~Gyr forward in time.
The current position and velocity of Pal 5 are given by \citet{oden03}. In particular the position of Pal 5 in the Galaxy is $(x, y, z) = (8.2, 0.2, 16.6)$~kpc; here $x$, $y$, and $z$ denote { righthanded} Galactocentric Cartesian coordinates, where the Sun has coordinates (-8.0, 0.0, 0.0). We assumed a radial velocity $v_r=-44.3$~km/s (observer at rest at the present location of the Sun), a tangential velocity (as seen by observer at rest at the present location of the Sun) $v_t=90$~km/s with a P.A.$=231^\circ$ with respect to the direction pointing to the northern equatorial pole, and $280^\circ$ with respect to Galactic north, as implied by the fit of the local orbit \citep{oden03}. These data yield a spatial velocity with components $(-40.7, -89.3, -21.0)$~km/s in the Galactic reference frame mentioned above.

\section{Results and discussion}\label{results}

\begin{figure}
\centering
\includegraphics[trim=4.5cm 10cm 2.5cm 11cm, clip=true, width=0.59\textwidth, angle=0]{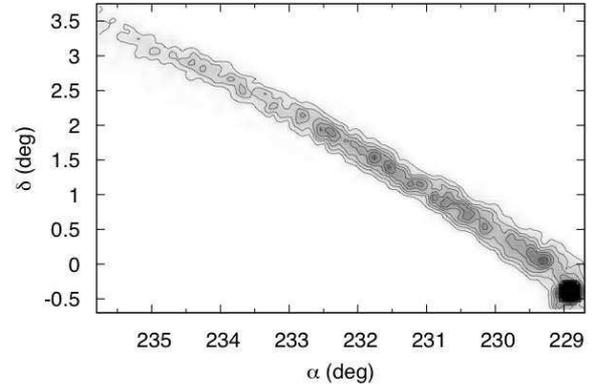}\\
\caption{The trailing  tail of the cluster in the restricted three-body model.
The contour lines in the tails refer to $470$, $940$, $1410$, $1970$, $2500$ and $3300$~M$_\odot/\rm{deg}^2$. }
\label{isod_mod}
\end{figure}

\begin{figure}
\centering
\includegraphics[width=0.41\textwidth]{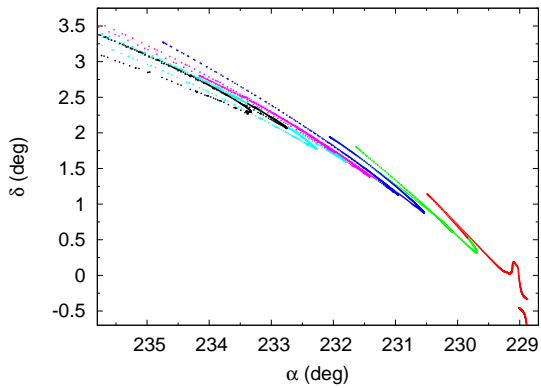}
\caption{Streaklines which show the positions, at the present time, of stars { that} have escaped from Pal 5 in the last $2.95$~Gyrs. Different colors correspond to stars lost between $0.$ and $0.5$~Gyr ago (red dots), between $0.5$ and $1.0$~Gyr ago (green dots), between $1.0$ and $1.5$~Gyr ago (blue dots), between $1.5$ and $2.0$~Gyr ago (magenta dots), between $2.0$ and $2.5$~Gyr ago (cyan dots), between $2.5$~Gyr and $2.95$~Gyr ago (black dots). }
\label{streak}
\end{figure}

After $2.95$~Gyr of evolution, $15$ disk crossings and $11$ pericenter passages, which every $\sim 270$~Myr bring the cluster { to} 
distances of $\sim5.9~$kpc from the Galaxy center,  the N-body simulation shows that the cluster has lost most of its initial mass ($88$\%), now redistributed { into} two long and narrow tidal tails.
The apparently smoothness of the tails  hides a  complex and  inhomogeneous stellar distribution that becomes evident only when plotting the isodensity contours, as in  Fig.~\ref{isoall} for the whole stellar stream. 
This plot reveals that indeed the tails are characterized by several substructures, and it shows that currently the densest  are all localized in the trailing tail at distances between 0.5$\degr$ and 3.5$\degr$  from the cluster center.
In this portion of the tail, our N-body simulation predicts { there are}
two prominent density enhancements (see top panel of Fig.~\ref{all}): the first located very close to the cluster ($\alpha<$ 230$\degr$) and the second starting at $\alpha$=230.5$\degr$  and ending at $\alpha$=232$\degr$. The second region shows an underdense region in it, whose extension is $\sim$ 0.5$\degr$.
All these characteristics, at the same location, are also found in the observed streams (cf, for example, Fig.~3, in \citealt{oden03}). 
Not only { does} the position of these clumps closely resemble the location of those observed in Pal 5's tails, but their linear densities (bottom panel of Fig. \ref{all}) -- 2--3 times above the density of the surrounding streams -- and the presence of underdense regions, with sizes of 0.5$^\circ$-1$^\circ$, also  agree with those measured for Pal 5 \citep[see Fig.~4,][]{oden03}.
These substructures are also  visible in the isodensity contours of the simplified numerical model (in Fig.~\ref{isod_mod}),  at few degrees  from the cluster center. 
When compared to the full N-body simulation,
the differences in the exact location and amplitude of the clumps in the restricted three-body problem { occur because} the former does not take the internal evolution of the cluster into account.
The gravitational potential of the cluster is kept fixed with time, and thus the instantaneous mass loss { that} affects the density of the innermost regions of the tails can only  be captured approximatively { unlike in} the N-body simulation, where these phenomena are modeled self-consistently.
This  ultimately leads to some differences in the density distribution along the streams, with clumps arising in the simplified numerical model { that are} less massive and less dense.
Nevertheless, both approaches unequivocally show substructures in models of Pal 5's tidal tails, a few degrees from the cluster center (see Appendix \ref{app2} for  the dependency of the substructures on the orbital parameters).  \\
These regions of over- and underdensities are due to the  epicyclic motion of star escaping from the cluster: stars lost at different times redistribute along the tail following a complex path, as shown by the streaklines  in  Fig.~\ref{streak}. 
This kinematic process, described and studied in a number of papers 
\citep{kupper08, just09, kupper10, kupper12, lane12}, is applied here for
the first time to reproduce observed stellar streams, including
both the position and the intensity of the observed stellar
inhomogeneities.
 Also the absence of  overdensities in the trailing tail at distances greater than 4$^\circ$--5$^\circ$ is simply { because} we have integrated Pal 5's orbit for the last $\sim$3 Gyr. With this choice, at the current cluster position, most of the escaped stars do not reach projected distances { over} 4$^\circ$--5$^\circ$ from the center of the cluster  \footnote{The length of tidal tails depends on the location of the GC along its orbit: for example, at the last pericenter passage, $160$Myr ago, our models show that stars were spread along a 12~kpc stream, 2.2 times longer than the extension of the simulated Pal 5's stream at the current position.}. If the orbit is integrated over much longer times ($\sim$8 Gyr), streaklines show that epicycle loops also form at distances of several kpc from the center. This process is thus able to form over- and underdensities in several regions of the tails, at a variety of distances from Pal 5.

 Moreover, that these substructures are also reproduced in models where the mutual gravity is neglected rules out Jeans instability as a possible mechanism responsible for their formation, as suggested by \citet{quillen10}. This hypothesis has been already questioned by  \citet{schneid11}, and here we confirm that it can be excluded as the mechanism at the origin of the substructures observed in Pal 5.

The association of overdense regions along the tails to the location of epicycles (upper panel of Fig.~\ref{all},  Figs~\ref{isod_mod}, and ~\ref{streak}) also rules out the possibility that they are due to an episodic mass loss from the cluster (see Appendix \ref{app1}
for further details).

\section{Conclusions}\label{conclusions}

We have studied the formation and characteristics of the tidal tails of the GC Pal 5 along its orbit in a smooth Milky Way potential by means of N-body simulations and simplified numerical models.
Our main results can be summarized as follows:
\begin{itemize}
\item The trailing tail of the cluster is more inhomogeneous than the leading tail.
\item The densest overdensities are found in the trailing tail at distances between 0.5$\degr$ and 3.5$\degr$  from the cluster center.
\item In this portion of the tail, our N-body simulation predicts { there are} two prominent density enhancements, the first located very close to the cluster ($\alpha<$ 230$\degr$) and the second starting at $\alpha$=230.5$\degr$  and ending at $\alpha$=232$\degr$. The second enhancement shows an underdense region, whose extension is $\sim$ 0.5$\degr$. These characteristics, at the same location, are found in the observed streams.
\item The linear density in the clumps is  two to three times above the density of the surrounding streams  and the underdense regions in between clumps have sizes of 0.5$^\circ$-1$^\circ$, comparable to those measured for Pal 5.
\item Pal 5's clumpy streams can also be formed in a smooth dark halo, { unlike in} previous findings  \citep{quinn08, yoon11},  as a simple consequence of the epicyclic motion of stars in the tails. 
\item Jeans instability and episodic mass loss can be ruled out as mechanisms at the origin of the observed substructures.
\end{itemize}

To conclude, our models can   reproduce the density contrast between the clumps and the surrounding tails found in the observed streams, without including any lumpiness in the dark matter halo.
At the same time, many works  \citep[among others, see][]{ibata02, johnston02} have shown that the structure, as well as the kinematics  \citep{SGV08},   of stellar streams can be very sensitive to heating by encounters with massive dark-matter subhalos. 
 It would be thus interesting to reconsider the impact of these substructures on the characteristics of Pal 5's stellar streams in light of these new results and, in particular, to set new upper limits on the granularity of the Milky Way dark halo.
Finally, future models may want to take also  inhomogeneities
in the Milky Way stellar disk into account (molecular clouds, spiral arms) to derive a
complete picture of the morphological evolution of stellar streams.

\section*{Acknowledgments}
This work was carried out under the HPC-EUROPA2 project (project number: 228398) with the support of the European Commission Capacities Area - Research Infrastructures Initiative. MM acknowledges financial support of the Observatoire de Paris.  
The authors wish to thank F. Combes, M. Bahi and D. Zidani for their permission to use the GPU-cluster ``Momentum'' and { their} help with it, and the anonymous referee for useful comments and suggestions.

\Online

\begin{appendix}

\section{Dependency on the orbital parameters} \label{app2}

\begin{figure}
\centering
\includegraphics[width =0.33\textwidth,angle=270]{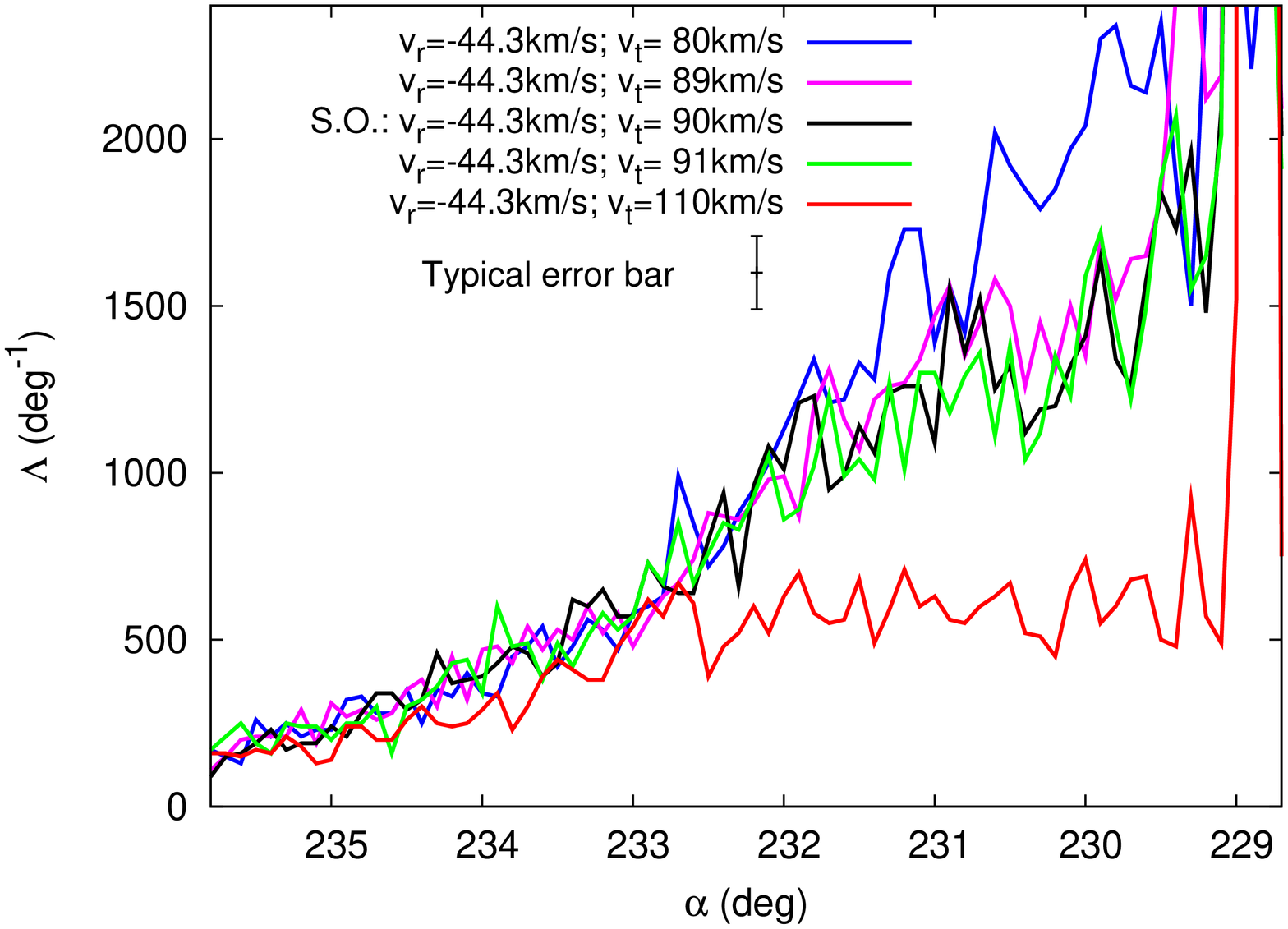}
\includegraphics[width =0.33\textwidth,angle=270]{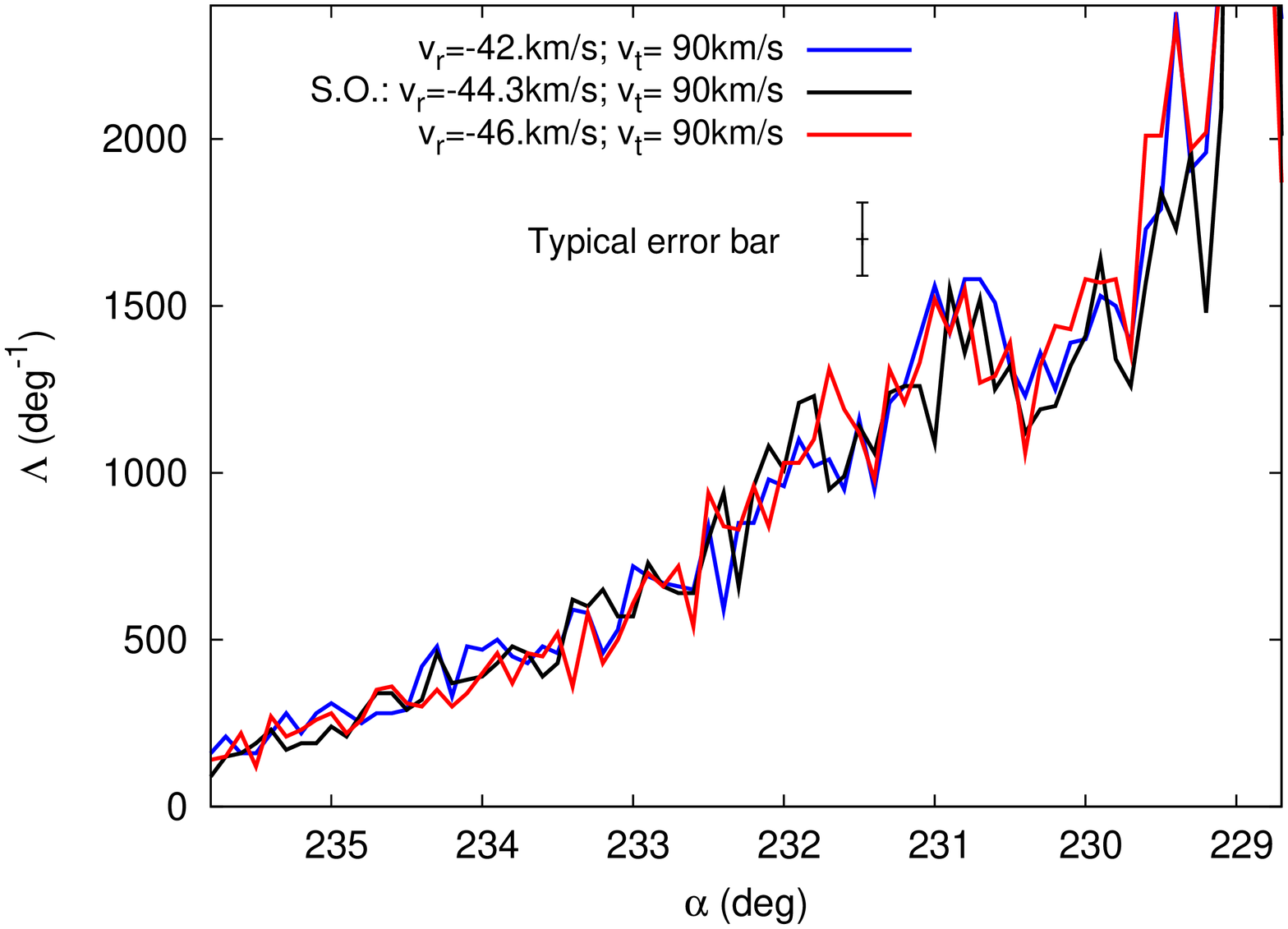}
\vspace{0.2cm}
\caption{Upper Panel: linear density of the trailing tail, as a
function of the distance from the cluster center, for different orbits
having the same radial velocity of the standard orbit (S.O., black line),
i.e.  $-44.3$~km/s,  and $v_t=80$~km/s (blue line), $v_t=89$~km/s (magenta line), $v_t=91$~km/s (green line), or $v_t=110$~km/s (red line). Lower Panel:  The same but for orbits having the same tangential
velocity of the standard orbit (S.O., black line), i.e. $90$~km/s, and $v_r=-42$~km/s (blue line) or $v_r=-46$~km/s (red line). The typical Poisson's uncertainty is shown.}
\label{vrvt}
\end{figure}

\begin{figure*}
\centering
\vspace{0.2cm}
\includegraphics[trim=5cm 7.cm 4.5cm 11cm, clip=true,  width =0.39\textwidth,angle=0]{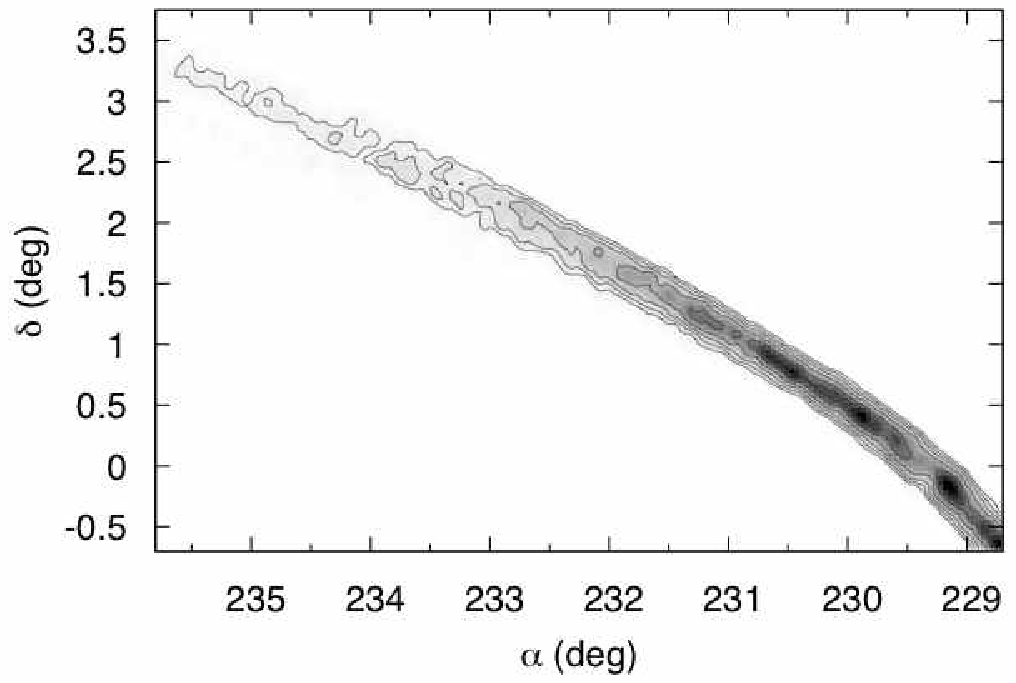}\hspace{-1cm}
\includegraphics[trim=5cm 7cm 4.5cm 11cm, clip=true, width = 0.39\textwidth,angle=0]{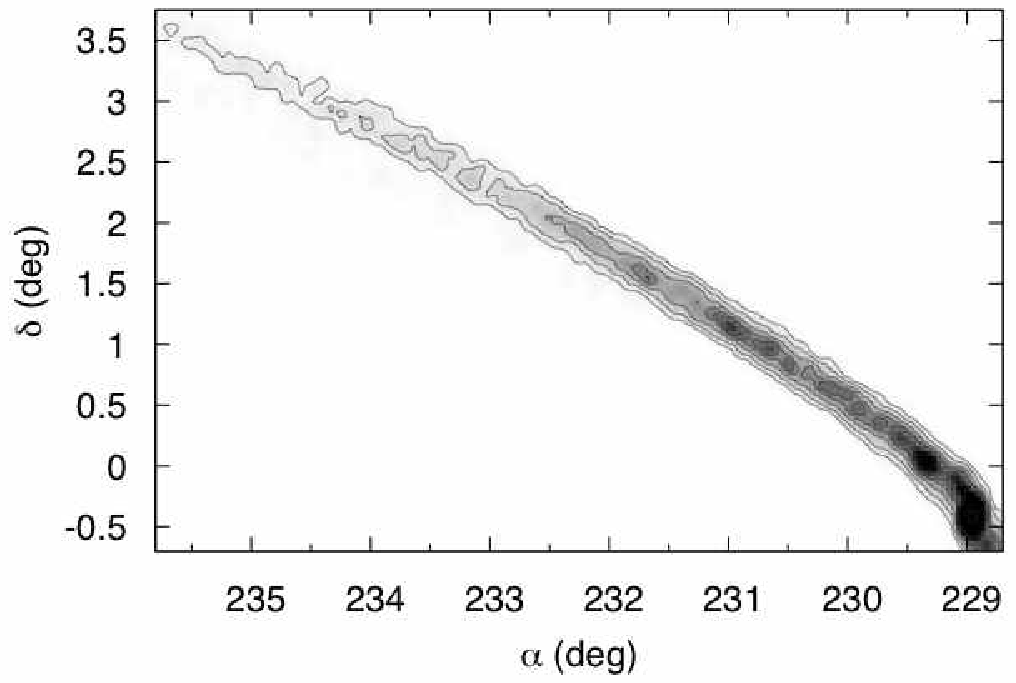}\\
\vspace{-1.5cm}
\includegraphics[trim=5cm 7.cm 4.5cm 11cm, clip=true, width = 0.39\textwidth,angle=0]{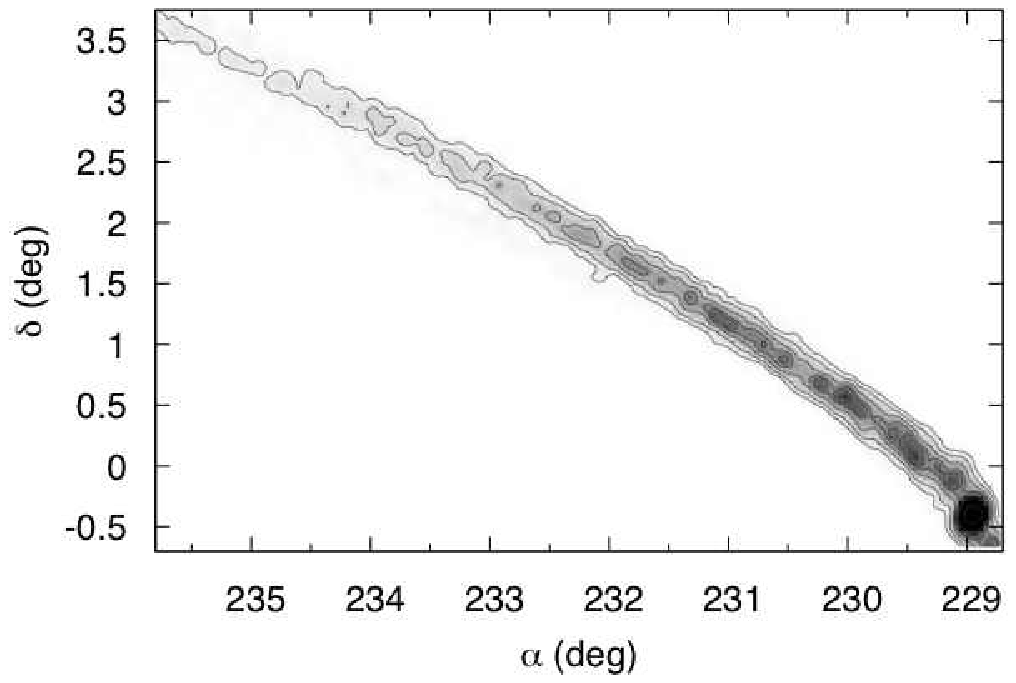}\hspace{-1cm}
\includegraphics[trim=5cm 7.cm 4.5cm 11cm, clip=true, width = 0.39\textwidth,angle=0]{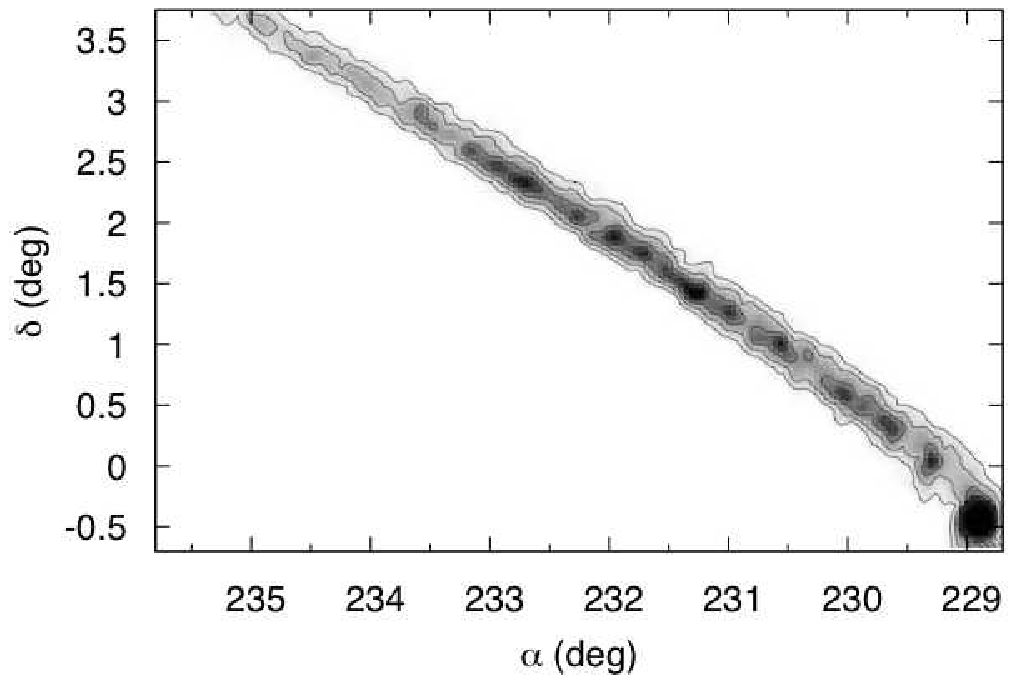}\\
\vspace{-1.5cm}
\includegraphics[trim=5cm 10.cm 4.5cm 11cm, clip=true, width = 0.39\textwidth,angle=0]{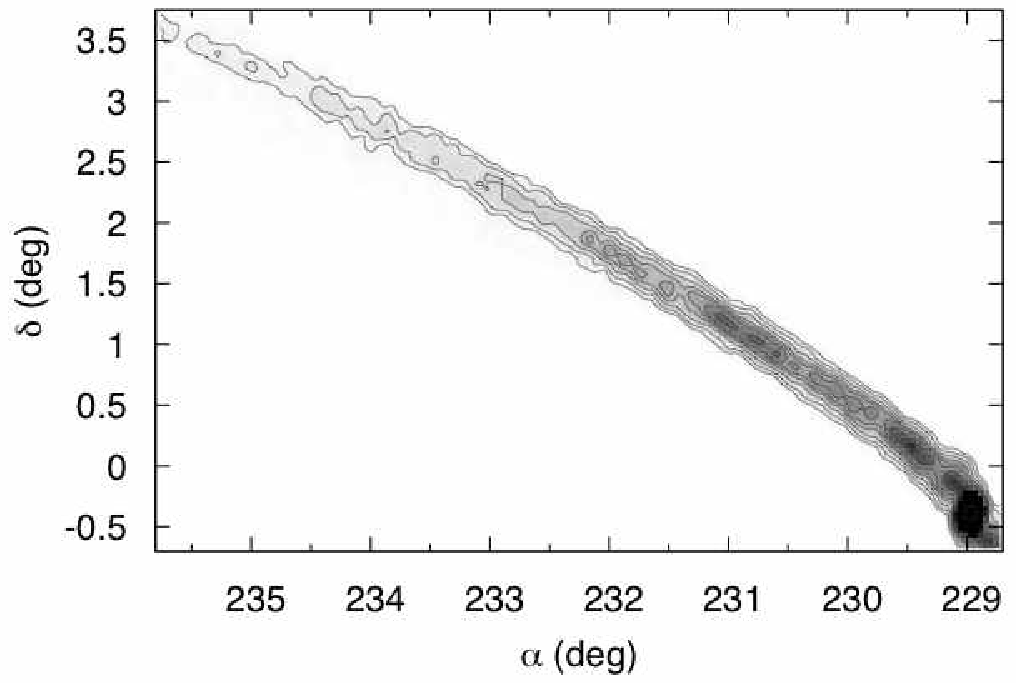}\hspace{-1cm}
\includegraphics[trim=5cm 10.cm 4.5cm 11cm, clip=true, width = 0.39\textwidth,angle=0]{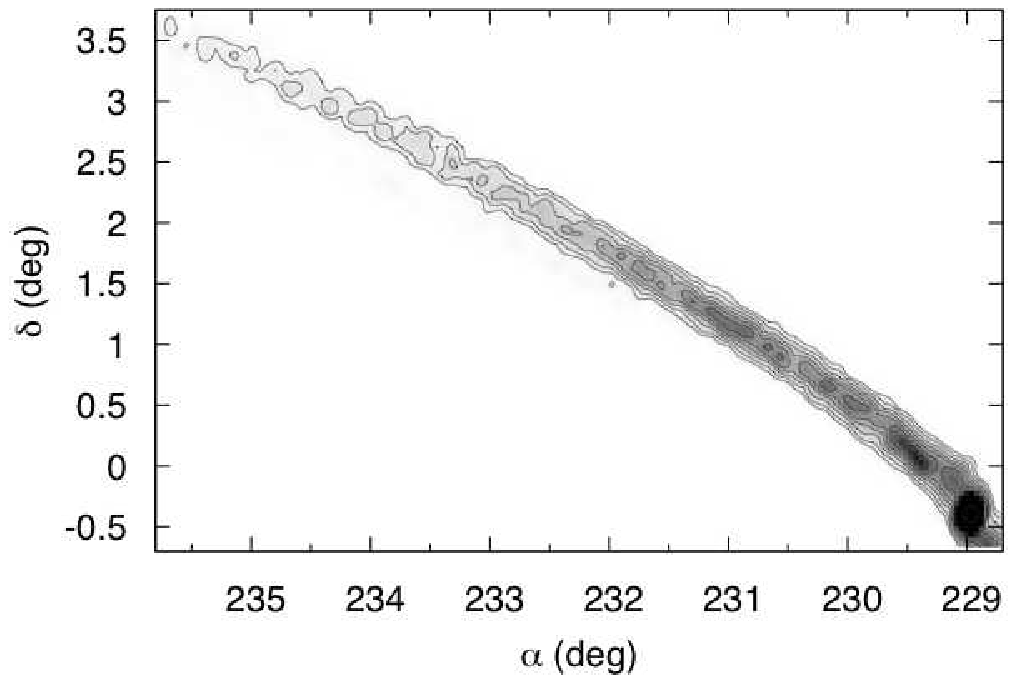}\caption{ Isodensity contour plots
for the trailing tail of the cluster for different N-body simulations having orbits
with  $v_r=-44.3$~km/s and $v_t=80$~km/s (upper left panel), $v_r=-44.3$~km/s and $v_t=89$~km/s (upper right panel), $v_r=-44.3$~km/s and $v_t=91$~km/s (middle left panel), $v_r=-44.3$~km/s and $v_t=110$~km/s (middle right panel), $v_r=-42$~km/s and $v_t=90$~km/s (bottom left panel), $v_r=-46$~km/s and $v_t=90$~km/s (bottom right panel). See upper panel of Fig. \ref{all} for a comparison with the isodensity contours
of the standard orbit}
\label{mapsvrvt}
\end{figure*}

In Section~\ref{results}, we have seen that epicyclic motion of stars in the tails generate overdensities mostly in the trailing tail of Pal 5, at distances of few degrees from the cluster center.  Here we show how sensitive this finding is to a change in the orbital parameters. To this aim, we have run several N-body simulations
changing both the radial, $v_r$, and tangential, $v_t$ velocities with respect to those of the standard orbit (we 
recall the reader that for the standard orbit, $v_r=-44.3$~km/s and $v_t=90$~km/s). In particular, for the tangential velocity, we adopted the extreme values, $80$~km/s and $110$~km/s, beyond which the orbit
fails to give a good representation of the observations, as shown by \citet{oden03}, and also 
two intermediate values similar to those of the standard orbit, $v_t=89$~km/s and $v_t=91$~km/s.
For the radial velocity we arbitrarily varied the best fit parameter by $\pm 2$~km/s, since the error on this value is
smaller than the one affecting the tangential velocity \citep{oden03}.
In all the models presented in this Section, the position angle P.A. has been kept fixed to $=231^\circ$  and the internal parameters of the cluster have not been changed.
In Figs.~\ref{vrvt} and \ref{mapsvrvt} the results of this analysis are presented.
The orbits with $v_r=-44.3$~km/s and $v_t=110$~km/s  or $v_t=80$~km/s are significantly different from 
the standard orbit. On the first orbit (which has a pericenter $\sim 1.2$~kpc greater than that of the standard orbit)
the cluster loses a small percentage of its mass, and the tails are characterized by a low density and a nearly flat stellar distribution, with density variations within the error bars. 
On the other hand, although the cluster with $v_r=-44.3$~km/s and $v_t=80$~km/s is destroyed by the tidal interaction with the Galactic field (the pericenter of the orbit is 4.7~kpc, a factor $\sim 1.3$ smaller than that of the standard orbit), clumps are clearly visible between $229^\circ$ and $231^\circ$.
Thus, for variation of the tangential velocity of the order of
10~km/s the density of the tails changes significantly.
For smaller changes of the tangential and radial velocities the mass loss rate is closer to that of the standard
orbit and stellar overdensities are present in the trailing tail, with similar positions and amplitudes.
This also confirms that the overdensities found for the standard orbit are not due 
to random fluctuations, but rather robust characteristics of this portion of the tail, since for small changes
around the standard values they are still present, and generate a profile that closely 
resembles that of the standard orbit.


\begin{figure*}[!h]
\centering
\includegraphics[width =0.33\textwidth,angle=270]{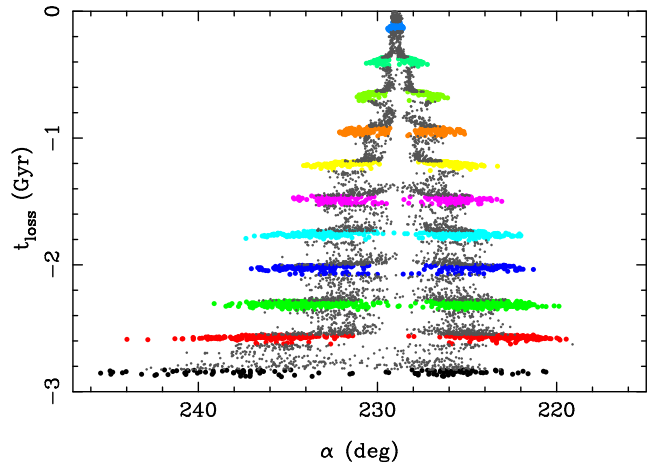}
\vspace{0.2cm}
\caption{Time, $t_{loss}$, at which escaped stars crossed the tidal boundary of the cluster as a function of the right ascension $\alpha$ of stars in Pal 5's tails. The right ascension has been evaluated at the current  epoch (see Fig.~\ref{isoall} for comparison). $t_{loss}$ is zero for stars lost from the GC at the current epoch, and attends the lowest values for stars escaped from the GC at the beginning of the simulation. Each color in the plot corresponds to stars lost at different pericenter passages, while grey points correspond to stars lost between two consecutive pericenter passages.}
\label{times}
\end{figure*}
\section{Why are clumps not caused by an episodic mass loss?} \label{app1}
In their modeling of Pal 5's tidal tails, \citet{dehnen04} have already shown that at any given distance from the cluster center, stars lost from the cluster at different times  can be found. We confirm this finding in  Fig.~\ref{times}, where we plot  the time stars escape from the tidal boundary  \footnote{For simplicity we adopted the current  tidal radius of the cluster as reference value.}  as a function of the right ascension $\alpha$, which describes the spatial extension of the streams.
At any given value of $\alpha$ corresponds many different values of $t_{loss}$: this is especially the case for particles in the inner part of the tails, at distances of $3\degr-4\degr$ from the cluster center. Moreover, the distribution in the plane $\alpha-t_{loss}$ resembles a tree, with branches and trunk corresponding, respectively, to stars lost at each  pericenter passage (plotted in Fig.~\ref{times} with different colors), and in between two consecutive pericenter passages. Due to their higher velocity dispersion when they escape the cluster,  stars lost at the pericenter passage are rapidly spread over a larger extension of the tidal tails than stars lost in between consecutive pericenter passages, which are systematically, for every  $t_{loss}$, closer to the cluster center than stars lost at pericenters. This behavior in the redistribution of stars lost in different phases of the cluster orbit clearly demonstrates that any temporary accumulation outside the tidal boundary of stars lost at the pericenter passage is rapidly cleaned out and thus cannot be a mechanism able to produce stellar overdensities in tidal streams at several degrees of distance from the cluster center.

\end{appendix}

\end{document}